\documentclass[fleqn,twoside]{article}
\usepackage{espcrc2,psfrag}

\usepackage{graphicx}
\usepackage[figuresright]{rotating}

\newcommand{\AmS}{{\protect\the\textfont2
  A\kern-.1667em\lower.5ex\hbox{M}\kern-.125emS}}

\title{Precision flavour physics with $B\to K\nu\bar\nu$
        and $B\to Kl^+l^-$}

\author{G. Buchalla\address[CERN]{CERN, Theory Division, CH-1211 Geneva 23,
        Switzerland\\ and\\  
     Ludwig-Maximilians-Universit\"at M\"unchen, Fakult\"at f\"ur Physik,\\
     Arnold Sommerfeld Center for Theoretical Physics, D-80333 M\"unchen,
     Germany}}

\begin{document}


\thispagestyle{empty}

\begin{flushright}
CERN-PH-TH/2010-232\\
October 2010
\end{flushright}

\vspace{3.0truecm}
\begin{center}
\boldmath
\large\bf Precision flavour physics with 
$B\to K\nu\bar\nu$ and $B\to Kl^+l^-$
\unboldmath
\end{center}

\vspace{0.9truecm}
\begin{center}
Gerhard Buchalla\\[0.1cm]
{\sl CERN, Theory Division, CH-1211 Geneva 23,
        Switzerland\\ and\\
     Ludwig-Maximilians-Universit\"at M\"unchen, Fakult\"at f\"ur Physik,\\
     Arnold Sommerfeld Center for Theoretical Physics, D-80333 M\"unchen,
     Germany}
\end{center}

\vspace{0.9truecm}

\begin{center}
{\bf Abstract}
\end{center}

{\small
\vspace{0.2cm}\noindent
We discuss how the combined analysis of $B\to K\nu\bar\nu$ and 
$B\to Kl^+l^-$ can provide us with new physics tests practically 
free of form factor uncertainties.
Residual theory errors are at the level of several percent. This study
underlines the excellent motivation for measuring these modes at
a Super Flavour Factory, or, in the case of $B\to Kl^+l^-$, also
at a hadron collider.}

\vspace{0.9truecm}

\begin{center}
{\sl Invited talk at the 3rd Workshop on Theory, Phenomenology and 
Experiments in Heavy Flavour Physics\\
Capri, Italy, 5--7 July 2010\\
To appear in the proceedings}
\end{center}


\newpage
\thispagestyle{empty}
\vbox{}
\newpage
 
\setcounter{page}{1}


\begin{abstract}
We discuss how the combined analysis of $B\to K\nu\bar\nu$ and $B\to Kl^+l^-$
can provide us with new physics tests practically free of form factor 
uncertainties.
Residual theory errors are at the level of several percent. This study
underlines the excellent motivation for measuring these modes at
a Super Flavour Factory, or, in the case of $B\to Kl^+l^-$, also 
at a hadron collider.
\vspace{1pc}
\end{abstract}

\maketitle

\mathindent=0pt

\section{INTRODUCTION}

One of the best opportunities for precision tests of flavour physics
will be provided by the
study of $b\to s\nu\bar\nu$ transitions, induced by interactions at very
short distances. 
A measurement of the inclusive decay $B\to X_s\nu\bar\nu$, ideal from a
theory perspective, appears to be extremely challenging experimentally.
More promising is the measurement of exclusive channels such as
$B\to K\nu\bar\nu$, $B\to K^*\nu\bar\nu$. In this case a clean theoretical
interpretation requires, however, the control of nonperturbative
hadronic form factors. Direct calculations of form factors suffer from
sizable uncertainties. 
These can be greatly reduced through a combined analysis of the
rare decays $B\to K\nu\bar\nu$ and $B\to K l^+l^-$.
This option allows us to construct precision observables for testing the
standard model and for investigating new physics effects.
In particular neither isospin nor $SU(3)$ flavour symmetry are required
and form factor uncertainties can be eliminated to a large extent.
A detailed study has recently been given in \cite{Bartsch:2009qp}, 
where further details can be found. Similar ideas had been considered 
independently in \cite{Hewett:2004tv}.

{}From experiment only upper limits are available
for the branching ratios of the neutrino modes:
\cite{Amsler:2008zzb,Barberio:2008fa,Chen:2007zk,Aubert:2004ws}
\begin{eqnarray}
B(B^-\to K^-\nu\bar\nu) &<& 14\cdot 10^{-6} \label{kmnunuexp}\\
B(\bar B^0\to\bar K^0\nu\bar\nu) &<& 160\cdot 10^{-6} \label{k0nunuexp}
\end{eqnarray}
The most accurate experimental results for $B\to Kl^+l^-$
are from Belle \cite{Wei:2009zv}. The extrapolated, non-resonant
branching fraction is measured to be
\begin{equation}\label{kllexp}
B(B\to K l^+l^-) = (0.48^{+0.05}_{-0.04}\pm 0.03)\cdot 10^{-6}
\end{equation}
consistent with results from BaBar \cite{Aubert:2008ps}.
The recent paper \cite{Wei:2009zv} also contains information
on the $q^2$-spectrum in terms of partial branching fractions
for six separate bins. 
Similar results from CDF were reported in \cite{Pueschel:2010rm}.

\section{\boldmath THEORY OVERVIEW}

\subsection{Dilepton-mass spectra and short-distance coefficients}

We define the kinematic quantities $s=q^2/m^2_B$
(where $q^2$ is the dilepton invariant mass squared), $r_K=m^2_K/m^2_B$,
and
\begin{equation}\label{lkdef}
\lambda_K(s)=1+r^2_K+s^2-2 r_K-2s-2 r_K s
\end{equation}
The differential branching fractions for $\bar B\to\bar K\nu\bar\nu$ and
$\bar B\to\bar Kl^+l^-$ can then be written as 
\begin{eqnarray}\label{dbknnds}
&&\hspace*{-0.6cm}\frac{dB(\bar B\to\bar K\nu\bar\nu)}{ds} =
\tau_B\frac{G^2_F\alpha^2m^5_B}{256\pi^5}
|V_{ts}V_{tb}|^2\cdot \nonumber\\
&& \lambda^{3/2}_K(s) f^2_+(s)\, |a(K\nu\nu)|^2
\end{eqnarray}
\begin{eqnarray}\label{dbkllds}
&&\hspace*{-0.6cm}\frac{dB(\bar B\to\bar Kl^+l^-)}{ds} =
\tau_B\frac{G^2_F\alpha^2m^5_B}{1536\pi^5}
|V_{ts}V_{tb}|^2\cdot \nonumber\\ 
&&  \lambda^{3/2}_K(s) f^2_+(s)\left(|a_9(Kll)|^2 +|a_{10}(Kll)|^2\right)
\end{eqnarray}
The coefficient $a(K\nu\nu)$ is given by
a short-distance Wilson coefficient at the weak scale, which is
known very precisely.
The coefficient $a_9(Kll)$ contains the Wilson
coefficient $\tilde C_9(\mu)$ combined with the short-distance
kernels of the $\bar B\to\bar Kl^+l^-$ matrix elements of four-quark
operators evaluated at $\mu={\cal O}(m_b)$. The coefficient $a_9(Kll)$
multiplies the local operator $(\bar sb)_{V-A}(\bar ll)_V$.
At next-to-leading order (NLO) the result can be extracted from the
expressions for the inclusive decay $\bar B\to X_sl^+l^-$
given in \cite{Buchalla:1995vs,Buras:1994dj,Misiak:1992bc},
where also the Wilson coefficients and operators of the
effective Hamiltonian and further details can be found.
The coefficient $a_{10}(Kll)$ is a short-distance quantity,
which is precisely known, similarly to $a(K\nu\nu)$.

\subsection{Form factors}

The long-distance hadronic dynamics of $\bar B\to\bar K\nu\bar\nu$ and
$\bar B\to\bar Kl^+l^-$ is contained in the matrix elements
\begin{eqnarray}
&&\hspace*{-0.6cm}\langle\bar K(p')|\bar s\gamma^\mu b|\bar B(p)\rangle
= f_+(s)\, (p+p')^\mu + \nonumber\\
&&[f_0(s)-f_+(s)]\,\frac{m^2_B-m^2_K}{q^2}q^\mu
\label{fpf0def}
\end{eqnarray}
\begin{eqnarray}
&&\hspace*{-0.6cm}\langle\bar K(p')|\bar s\sigma^{\mu\nu}b|\bar B(p)\rangle
=\nonumber\\
&& i\frac{f_T(s)}{m_B+m_K}\left[(p+p')^\mu q^\nu - q^\mu (p+p')^\nu\right]
\label{ftdef}
\end{eqnarray}
which are parametrized by the form factors $f_+$, $f_0$ and $f_T$.
Here $q=p-p'$ and $s=q^2/m^2_B$. The term proportional to $q^\mu$ in
(\ref{fpf0def}), and hence $f_0$, drops out when the small lepton masses
are neglected as has been done in (\ref{dbknnds}) and (\ref{dbkllds}).
The ratio $f_T/f_+$ is independent of unknown hadronic
quantities in the small-$s$ region due to the relations
between form factors that hold in the limit of large kaon energy
\cite{Charles:1998dr,Beneke:2000wa}
\begin{equation}\label{ftfp}
\frac{f_T(s)}{f_+(s)}=\frac{m_B+m_K}{m_B}+{\cal O}(\alpha_s,\Lambda/m_b)
\end{equation}
Here we have kept the kinematical dependence on $m_K$ in the
asymptotic result.
In contrast to $f_+$ the form factor $f_T$ is scale and scheme dependent.
This dependence is of order $\alpha_s$ and has been neglected in (\ref{ftfp}).

We remark that the same result for $f_T/f_+$ is also obtained in the
opposite limit where the final state kaon is soft, that is in the region
of large $s={\cal O}(1)$ 
\cite{Wise:1992hn,Burdman:1992gh,Falk:1993fr,Casalbuoni:1996pg,Buchalla:1998mt}.
From this observation we expect (\ref{ftfp}) to be
a reasonable approximation in the entire physical domain. This is indeed
borne out by a detailed analysis of QCD sum rules on the light cone
\cite{Ball:2004ye}, which cover a range in $s$ from $0$ to $0.5$.

The impact of the $f_T/f_+$ term is numerically small,
about $13\%$ of the amplitude $a_9(Kll)$.
A $15\%$ uncertainty, which may be expected for the
approximate result (\ref{ftfp}), will only imply an uncertainty of $2\%$
for $a_9(Kll)$ or the $\bar B\to\bar K l^+l^-$ differential rate.
In practice, this leaves us with the form factor $f_+(s)$ as the essential
hadronic quantity for both $\bar B\to\bar K\nu\bar\nu$ and
$\bar B\to\bar K l^+l^-$.

\begin{figure}[t]
\begin{center}
\resizebox{8cm}{!}{\includegraphics{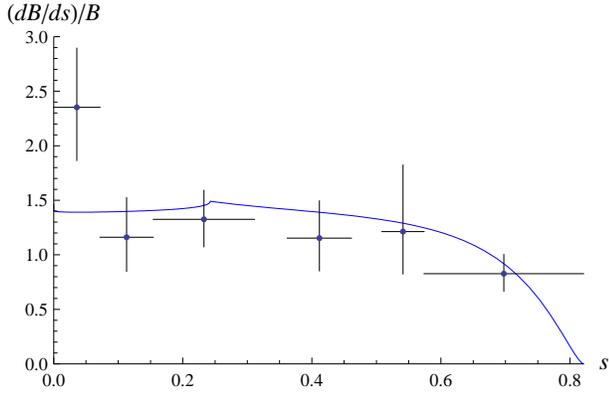}}
\caption{\label{fig:bestfit}
The shape of the $\bar B\to\bar K l^+l^-$ spectrum, $(dB/ds)/B$,
from the Belle data \cite{Wei:2009zv} (crosses)
and from theory with the best-fit shape parameters
$a_0=1.6$, $b_1/b_0=1$ (solid curve).}
\end{center}
\end{figure}

We employ the parametrization proposed by Becirevic and Kaidalov
\cite{Becirevic:1999kt} in the form
\begin{equation}\label{ffparam}
f_+(s)\equiv f_+(0)\,\frac{1-(b_0+b_1-a_0 b_0)s}{(1-b_0 s)(1-b_1 s)}
\end{equation}
The parameter $b_0$ is given by
\begin{equation}\label{b0def}
b_0=\frac{m^2_B}{m^2_{B^*_s}}\approx 0.95 \qquad {\rm for} \quad
m_{B^*_s}=5.41\,{\rm GeV}
\end{equation}
$b_0$ represents the position of the $B^*_s$ pole and is
taken as fixed, following \cite{Becirevic:1999kt}.
The remaining three quantities $a_0$, $b_1$ and $f_+(0)$ are treated
as variable parameters. QCD sum rules on the light cone
(LCSR) give \cite{Ball:2004ye,Bartsch:2009qp}
\begin{equation}\label{a0b1f0}
f_+(0)=0.304\pm 0.042,\quad a_0\approx 1.5, \quad b_1=b_0
\end{equation}
$f_+(0)$ only affects the overall normalization of the decay
rates and cancels in appropriate ratios.
Combining theoretical constraints \cite{Bartsch:2009qp}, we adopt the 
following default ranges for the shape parameters
\begin{equation}\label{a0b1range}
1.4\leq a_0\leq 1.8\qquad\quad  0.5\leq b_1/b_0\leq 1.0
\end{equation}
with
\begin{equation}\label{a0b1ref}
a_0=1.6\qquad\quad  b_1/b_0=1.0
\end{equation}
as our reference values.
The latter are also obtained \cite{Bartsch:2009qp} as the best fit to 
the shape of the measured $\bar B\to\bar K l^+l^-$ spectrum in 
Fig. \ref{fig:bestfit}. 

\subsection{\boldmath $\bar B\to \bar K l^+l^-$: Nonperturbative corrections}

In this section we comment on the theoretical framework for
$\bar B\to \bar K l^+l^-$ and on nonperturbative effects beyond those
that are contained in the form factors.

It is well known that, because of huge backgrounds from
$\bar B\to\bar K\psi^{(')}\to\bar Kl^+l^-$, the region of $q^2$ containing
the two narrow charmonium states $\psi=\psi(1S)$ and $\psi'=\psi(2S)$
has to be removed by
experimental cuts from the $q^2$ spectrum of $\bar B\to \bar K l^+l^-$.
The overwhelming background from $\psi$ and $\psi'$ is related to a drastic
failure of quark-hadron duality in the narrow-resonance region for the
{\it square\/} of the charm-loop amplitude, as has been discussed
in \cite{Beneke:2009az}.
Nevertheless, the parts of the $q^2$ spectrum below and above the
narrow-resonance region remain under theoretical control and are sensitive
to the flavour physics at short distances.
A key observation here is that the amplitude is largely dominated by the
semileptonic operators
\begin{eqnarray}\label{q9q10}
Q_9 &=& (\bar sb)_{V-A}(\bar ll)_V \nonumber\\
Q_{10} &=& (\bar sb)_{V-A}(\bar ll)_A
\end{eqnarray}
which have large coefficients $\tilde C_9$ and $\tilde C_{10}$.
These contributions are perturbatively calculable up to the long-distance
physics contained in the form factor $f_+(s)$.
The $\bar B\to\bar Kl^+l^-$ matrix elements of four-quark operators,
such as $(\bar sb)_{V-A}(\bar cc)_{V-A}$, are more complicated, but still
systematically calculable.
Schematically, the $\bar B\to\bar Kl^+l^-$ rate is proportional to
\begin{equation}\label{c94qc10}
|\tilde C_9+\Delta_{4q}|^2 + |\tilde C_ {10}|^2
\end{equation}
where $\Delta_{4q}$ represents contributions from four-quark operators,
for instance charm loops or weak annihilation 
effects\footnote{Weak annihilation contributions to 
$\bar B\to\bar K l^+l^-$ are negligibly small, although they are a 
leading-power effect \cite{Bartsch:2009qp}.}.
Because $\Delta_{4q}$ is numerically subleading
(${\cal O}(10\%)$ of the total rate), the impact of any uncertainties
in its evaluation will be suppressed.

In the {\it low-$q^2$ region\/} $\Delta_{4q}$ can be computed using
QCD factorization \cite{Beneke:2001at}.
This approach, which is based on the heavy-quark limit and the large energy
of the recoiling kaon, should work well for the real part of the amplitude
in view of the experience from two-body hadronic $B$ decays
\cite{Beneke:2007zz} and $B\to K^*\gamma$ \cite{Bosch:2004nd}.

In the {\it high-$q^2$ region\/} the appropriate theoretical
framework for the computation of $\Delta_{4q}$ is an
operator product expansion exploiting the presence of the large
scale $q^2\sim m^2_b$.
This concept has been used in \cite{Buchalla:1998mt} in analyzing
the endpoint region of $b\to sl^+l^-$, which is governed by few-body
exclusive modes. A detailed treatment, including the discussion
of subleading corrections, has been given in \cite{Grinstein:2004vb}.
Power corrections are generally smaller than for low $q^2$.
Uncertainties could still come from violations of local quark-hadron
duality. The relative amplitude of oscillations
in $\tilde C_9 + {\rm Re}\,\Delta_{4q}$ may be estimated to be
of order $10$ to $20\%$.
We expect these local variations to be averaged out when the spectrum
is integrated over $s$ \cite{Buchalla:1998mt} such that the residual
uncertainty will be reduced. As discussed in \cite{Beneke:2009az},
global duality in this sense cannot be expected to hold for the
second order term $|\Delta_{4q}|^2$ in (\ref{c94qc10}).
On the other hand, this contribution is numerically very small,
at the level of few percent, and duality violations will only have
a minor effect.

\subsection{\boldmath $B^-\to\tau^-\bar\nu_\tau\to K^-\nu_\tau\bar\nu_\tau$}
\label{subsec:nunubkgr}

The decay  $B^-\to\tau^-\bar\nu_\tau$ followed by $\tau^-\to K^-\nu_\tau$
produces a background for the short-distance reaction $B^-\to K^-\nu\bar\nu$
at the level of $15$-$25\%$,
which has been discussed recently in \cite{Kamenik:2009kc}.
This background has to be taken into account
for a precise measurement of the short-distance branching fraction
$B(B^-\to K^-\nu\bar\nu)$. It needs to be subtracted from the
experimental signal, but this should ultimately be possible with
essentially negligible uncertainty \cite{Bartsch:2009qp}.

\section{PRECISION OBSERVABLES}
\label{sec:precision}

Our predictions for the branching fractions in the standard model are
\begin{eqnarray}\label{bnnnum}
&& \hspace*{-0.6cm}B(B^-\to K^-\nu\bar\nu)\cdot 10^6  = 
4.4\, ^{+1.3}_{-1.1}\, (f_+(0))\nonumber\\
&&\,\, ^{+0.8}_{-0.7}\, (a_0)\,\, ^{+0.0}_{-0.7}\, (b_1)
\end{eqnarray}
\begin{eqnarray}\label{bllnum}
&& \hspace*{-0.6cm}B(B^-\to K^-l^+l^-)\cdot 10^6 = 
0.58\, ^{+0.17}_{-0.15}\, (f_+(0))\nonumber\\
&&\,\, ^{+0.10}_{-0.09}\, (a_0)\,\, ^{+0.00}_{-0.09}\, (b_1)
\,\, ^{+0.04}_{-0.03}\, (\mu)
\end{eqnarray}

Whereas the individual branching fractions (\ref{bnnnum}) and
(\ref{bllnum}) suffer from large hadronic uncertainties, we expect their
ratio to be under much better theoretical control. It is obvious that the
form factor normalization $f_+(0)$ cancels in this ratio.
Moreover, the shape of the $q^2$ spectrum is almost identical for the
two modes. This is because the additional $q^2$-dependence
from charm loops in $B\to Kl^+l^-$, compared to $B\to K\nu\bar\nu$,
is numerically only a small effect outside the region of the
narrow charmonium states.
As a consequence, also the dependence on the form factor shape
will be greatly reduced in the ratio
\begin{equation}\label{rdef}
R=\frac{B(B^-\to K^-\nu\bar\nu)}{B(B^-\to K^-l^+l^-)}
\end{equation}
Numerically we find
\begin{equation}\label{rnum}
R=7.59\, ^{+0.01}_{-0.01}\, (a_0)\,\, ^{+0.00}_{-0.02}\, (b_1)
\,\, ^{-0.48}_{+0.41}\, (\mu)
\end{equation}
This prediction is independent of form factor uncertainties
for all practical purposes. It is limited essentially by
the perturbative uncertainty at NLO of $\pm 6\%$.
Using the experimental result in (\ref{kllexp}), the theory
prediction (\ref{rnum}), and assuming the validity of the
standard model, we obtain
\begin{eqnarray}\label{knnrkll}
&& \hspace*{-0.6cm}B(B^-\to K^-\nu\bar\nu) = 
R\cdot B(B^-\to K^-l^+l^-)_{exp} =\nonumber\\
&&(3.64\pm 0.47)\cdot 10^{-6}
\end{eqnarray}
With an accuracy of $\pm 13\%$, limited at present by
the experimental error, this result is currently the most
precise estimate of $B(B^-\to K^-\nu\bar\nu)$.

In order to obtain theoretically clean observables, the
region of the two narrow charmonium resonances $\psi(1S)$ and
$\psi(2S)$ has to be removed from the $q^2$ spectrum of $B\to Kl^+l^-$.
This leaves two regions of interest, the low-$s$ region below
the resonances, and the high-$s$ region above.
For the present analysis we define these ranges as
\begin{equation}\label{slohi}
\begin{array}{rl}
{\rm low}\ s: \qquad\quad & 0\leq s \leq 0.25\\
{\rm high}\ s: \qquad\quad & 0.6\leq s \leq s_m
\end{array}
\end{equation}
The resonance region $0.25 < s < 0.6$ corresponds to
the $q^2$ range $7\,{\rm GeV}^2 < q^2 < 16.7\,{\rm GeV}^2$.
For our standard parameter set the total rate for
$B\to K\nu\bar\nu$ or $B\to Kl^+l^-$ (non-resonant)
is divided among the three regions, low-$s$, narrow-resonance, high-$s$,
as $35 : 48 : 17$.

We first concentrate on the low-$s$ region, where $B^-\to K^- l^+l^-$
can be reliably calculated. To ensure an optimal cancellation of the
form factor dependence, one may restrict also the neutrino mode
to the same range in $s$ and define
\begin{equation}\label{r25def}
R_{25}\equiv \frac{\int_0^{0.25}ds\ dB(B^-\to K^-\nu\bar\nu)/ds}{
\int_0^{0.25}ds\ dB(B^-\to K^- l^+l^-)/ds}
\end{equation}
This ratio is determined by theory to very high precision.
Displaying the sensitivity to the shape parameters and the
renormalization scale one finds
\begin{equation}\label{r25num}
R_{25}=7.60\, ^{-0.00}_{+0.00}\, (a_0)\,\, ^{-0.00}_{+0.00}\, (b_1)
\,\, ^{-0.43}_{+0.36}\, (\mu)
\end{equation}
The form factor dependence is seen to cancel almost perfectly in $R_{25}$.
The shape pa\-ra\-me\-ters affect this quantity at a level of only 0.5 per
mille. One is therefore left with the perturbative uncertainty, estimated
here at about $\pm 5\%$ at NLO.

The independence of any form factor uncertainties in $R_{25}$ comes at
the price of using only $35\%$ of the full $B^-\to K^-\nu\bar\nu$ rate.
We therefore consider a different ratio, which is defined by
\begin{equation}\label{r256def}
R_{256}\equiv \frac{\int_0^{s_m}ds\, dB_\nu/ds}{
\int_0^{0.25}ds\, dB_l/ds +
\int_{0.6}^{s_m}ds\, dB_l/ds}
\end{equation}
In this ratio the fully integrated rate of
$B^-\to K^-\nu\bar\nu$ is divided by the integrated rate
of $B^-\to K^- l^+l^-$ with only the narrow-resonance region removed.
This ensures use of the maximal statistics in both channels.
Due to the missing region in $B^-\to K^- l^+l^-$ the dependence on the
form factor shape will no longer be eliminated completely,
but we still expect a reduced dependence.
Numerically we obtain, using the same input as before,
\begin{equation}\label{r256num}
R_{256}=14.60\, ^{+0.28}_{-0.38}\, (a_0)\,\, ^{+0.10}_{-0.02}\, (b_1)
\,\, ^{-0.80}_{+0.62}\, (\mu)
\end{equation}
This estimate shows that the uncertainty from $a_0$ and $b_1$ is indeed
very small, at a level of about $\pm 3\%$. With better empirical
information on the shape of the spectrum this could be further improved.

We conclude that ratios such as those in (\ref{r25def}) and (\ref{r256def}),
or similar quantities with modified cuts, are theoretically very well under
control. They are therefore ideally suited for testing the standard model
with high precision.

\section{NEW PHYSICS}
\label{sec:newphysics}

The branching fractions of $B\to K\nu\bar\nu$ and $B\to Kl^+l^-$ are
sensitive to physics beyond the standard model.
In general, nonstandard dynamics will have a different impact on
$B\to K\nu\bar\nu$ and $B\to Kl^+l^-$. The excellent theoretical control
over the ratios $R_{25}$ or $R_{256}$ will help to reveal even moderate
deviations from standard model expectations.

One example is the scenario with modified $Z$-penguin contributions
\cite{Buchalla:2000sk}, if these contributions
interfere destructively with those of the standard model.
In that case the ratios $R_{25}$ or $R_{256}$ could be
significantly suppressed. The modified $Z$-penguin scenario
may be realized, for instance, in supersymmetric models
\cite{Buchalla:2000sk,Altmannshofer:2009ma}.

Another class of theories that do change the ratios are those where
$B\to Kl^+l^-$ remains standard model like while $B\to K\nu\bar\nu$
receives an enhancement (or a suppression). Substantial enhancements of
$B(B\to K\nu\bar\nu)$ are still allowed by experiment, in fact much more
than for $B\to Kl^+l^-$.

A first example are scenarios with light invisible scalars $S$
contributing to $B\to KSS$, which has been suggested in
\cite{Bird:2004ts,Bird:2006jd} as an efficient probe of
light dark matter particles. $B\to KSS$ is also discussed
in \cite{Altmannshofer:2009ma}.
This channel adds to $B\to K\nu\bar\nu$, which is measured as
$B\to K + {\rm invisible}$. If the scalars have nonzero mass,
$B\to KSS$ could be distinguished from $B\to K\nu\bar\nu$ through the
missing-mass spectrum. On the other hand, if the mass of $S$ is small,
or the resolution of the spectrum is not good enough, a discrimination
of the channels may be difficult. The corresponding increase in
$B(B\to K\nu\bar\nu)$ could be cleanly identified through the ratios
$R_{25}$ and $R_{256}$.
Similar comments apply to the case where the invisible particles
are light (or massless) neutralinos $\tilde\chi^0_1$, which are
still allowed in the MSSM \cite{Dreiner:2009ic}. Substantial enhancements
of $B\to K + {\rm invisible}$ over the standard model expectation
through $B\to K\tilde\chi^0_1\tilde\chi^0_1$ are possible in the MSSM with
non-minimal flavour violation \cite{Dreiner:2009er}.

A further example is given by topcolor assisted technicolor
\cite{Buchalla:1995dp}. A typical scenario involves new strong dynamics,
together with extra $Z'$ bosons, which distinguishes the third generation
from the remaining two. The resulting flavour-changing neutral currents
at tree level may then predominantly lead to transitions between
third-generation fermions such as $b\to s\nu_\tau\bar\nu_\tau$.
An enhancement of $B(B\to K\nu\bar\nu)$ would result and might
in principle saturate the experimental bound (\ref{kmnunuexp}).
An enhancement of $20\%$, which should still be detectable, would probe
a $Z'$-boson mass of typically $M_{Z'}\approx 3\,{\rm TeV}$.
A similar pattern of enhanced $B\to K\nu\bar\nu$ and SM like
$B\to Kl^+l^-$ is also possible in generic $Z'$ models
\cite{Altmannshofer:2009ma}.

The subject of new physics in $b\to s\nu\bar\nu$ transitions has
been discussed in \cite{Grossman:1995gt} and more recently
in \cite{Altmannshofer:2009ma,Bird:2004ts,Bird:2006jd,Dreiner:2009er}.
New physics in $B\to K l^+l^-$ has been studied in \cite{Bobeth:2007dw}.

\section{CONCLUSIONS}
\label{sec:conclusion}

The strategy discussed here
puts $B\to K\nu\bar\nu$ as a new physics probe in the same class
as $K\to\pi\nu\bar\nu$, the `golden modes' of kaon physics.
Suitable ratios of (partially integrated) $B\to K\nu\bar\nu$ and 
$B\to Kl^+l^-$ decay rates are essentially free of form factor
uncertainties, while retaining sensitivity to interesting
New Physics scenarios.
$B\to K\nu\bar\nu$ together with $B\to Kl^+l^-$ thus hold
exciting opportunities for $B$ physics in the era of 
Super Flavour Factories.

{\bf Acknowledgements:}
I thank M.~Bartsch, M.~Beylich and D.-N.~Gao for their collaboration
on the topics presented in this talk, and the organizers of
Capri 2010 for a beautiful conference. The hospitality of the CERN Theory
Division is gratefully acknowledged.
This work was supported in part by the DFG cluster of excellence
`Origin and Structure of the Universe'.

\end{document}